# ARTICLE

# Electron transfer mediated decay of alkali dimers attached to He nanodroplets

L. Ben Ltaief†a, M. Shcherbinina, S. Mandalb, S. R. Krishnanc, R. Richterd, T. Pfeifere, M. Bauerf, A. Ghoshf, M. Mudrich†a, c, K. Gokhbergf, and A. C. LaForge†g

Alkali metal dimers attached to the surface of helium nanodroplets are found to be efficiently doubly ionized by electron transfer-mediated decay (ETMD) when photoionizing the helium droplets. This process is evidenced by detecting in coincidence two energetic ions created by Coulomb explosion and one low-kinetic energy electron. The kinetic energy spectra of ions and electrons are reproduced by simple model calculations based on diatomic potential energy curves, and are in agreement with *ab initio* calculations for the He-Na$_2$ and He-KRb systems. This work demonstrates that ETMD is an important decay channel in heterogeneous nanosystems exposed to ionizing radiation.

## Introduction

Upon electronic excitation or ionization of a weakly bound system, e.g. van-der-Waals clusters or hydrogen bonded complexes, by energetic photons, a highly non-equilibrium configuration of the electronic and nuclear degrees of freedom is prepared at the site of the photon's impact. In the subsequent relaxation of these systems various ultrafast interatomic/intermolecular energy and charge transfer processes involving neighboring sites become important. These processes currently attract considerable interest, because of their potential relevance for radiation damage of biological matter [1-3]. Among these relaxation mechanisms are interatomic/molecular Coulombic decay (ICD) [4] and electron transfer mediated decay (ETMD) [5]. ICD and ETMD lead, in addition to redistributing energy and charge throughout an extended system, to emission of low-kinetic energy electrons which are genotoxic and can induce irreparable damage in biological matter such as DNA-double strand breaks [1, 2]. Both processes become highly efficient when the local electronic decay by Auger process is energetically forbidden. They are present in many weakly-bound systems ranging from noble gas dimers to hydrogen-bonded molecular clusters [6, 7]. Recently, ICD and ETMD were observed in a hydrated biomolecule [8] and in aqueous media [9, 10], respectively.

ICD occurs by transfer of the energy stored in the initially perturbed site to a neighboring atom or molecule that leads to its ionization. ETMD, by contrast, proceeds by electron transfer from a donor to the initially created ion, such that the released energy leads to emission of a second electron either from the donor, ETMD(2), or from a second neighboring atom or molecule, ETMD(3), [5, 11, 12, 13]. It was shown to be the leading decay pathway if ICD is energetically forbidden [3, 14, 15]. In contrast to ICD where the final charge state of the electronically excited species remains constant, in an ETMD step its charge decreases by one electron charge. ETMD is also shown to be a much faster and stronger decay channel than its counterpart, radiative charge transfer (RCT) [14]. In RCT, the energy released upon neutralization of the ion is emitted as a photon [16-19] while in ETMD it is converted into kinetic energy of an emitted slow electron and of ionic fragments.

Helium (He) has a simple electronic structure and the highest ionization energy amongst all elements. Therefore, pure or doped He nanodroplets offer a unique medium where interatomic/molecular decays can be studied [14, 15, 20-29]. Owing to its chemical inertness and low temperature, attached dopant molecules are only weakly perturbed and tend to aggregate into weakly bound complexes inside the droplet or at the droplet surface. Recently, it was found that ETMD between a $He^+$ or a $He_2^+$ and a dopant attached to the droplet leads to the double ionization of the dopant [14, 15]. In particular, experiments in He nanodroplets doped with magnesium (Mg) clusters demonstrated that such a single photon double ionization process mediated by ETMD is highly efficient [15]. Unlike Mg clusters which are embedded inside the droplet, alkali dimers are adsorbed on its surface [25]. They can be formed either in the covalently bound ground $^1\Sigma$ state or in the lowest van der Waals $^3\Sigma$ state. The two states have markedly different electronic structures which result in different bond lengths, bond strengths, and dimer's orientation relative to the droplet surface. Moreover, since the

a. Department of Physics and Astronomy, Aarhus University, 8000 Aarhus C, Denmark
b. Indian Institute of Science Education and Research, Pune 411008, India
c. Indian Institute of Technology Madras, Chennai 600036, India
d. Elettra-Sincrotrone Trieste, 34149 Basovizza, Italy
e. Max-Plank-Institute für Kernphysik, 69117 Heidelberg, Germany
f. Physikalisch-Chemisches Institut, Universität Heidelberg, 69120 Heidelberg, Germany
g. Department of Physics, University of Connecticut, Storrs, Connecticut, 06269 USA

† ltaief@phys.au.dk, aaron.laforge@uconn.edu and mudrich@phys.au.dk

ETMD rate becomes large only at short distances between the $He^+(He_2^+)$ and the dopant [30], it will be accompanied by nuclear dynamics which involves both the ion and the alkali dimer. The dynamics should also reflect the binding properties and ultimately the electronic structure of the alkali dimer. Therefore, the latter might be imprinted on and can be observed in the electron and ion spectra [30].

In this paper we report the experimental observation of ETMD of different alkali metal dimers, $Na_2$, $K_2$, $Rb_2$, NaK, NaRb and KRb, attached to the surface of He nanodroplets. Contrary to the general concept of resonant charge migration towards the interior of the He droplet, which preferentially leads to charge transfer-induced ionization of dopants solvated inside the droplet, here we find that surface-bound alkali dimers and clusters are efficiently ionized by ETMD as well. In this work, this process is systematically investigated by measuring and simulating electron and ion kinetic energy distributions. Although alkali dimers attached to He droplet are a model system, this decay mechanism is relevant for any molecular complex where the ionization energy of one constituent exceeds the double ionization energy of another.

## Experiment

The experiments were performed at the GasPhase beamline of Elettra-Sincrotrone, Trieste, Italy. The experimental setup consists of a He nanodroplets source and a doping unit, a time-of-flight (TOF) spectrometer, and a velocity map imaging (VMI) spectrometer. A detailed description of the experimental setup can be found in Ref. [25].

Similar experimental conditions as in [29] are used in this work to generate, by continuous supersonic expansion of He, a beam of He nanodroplets with a mean droplet size of about 2000 He atoms per droplet, prior to doping [31]. The He nanodroplets are doped with alkali atoms (Rb, Na, K) in one or two heated cells with a length of 10 mm each. The doping level of metals is adjusted by setting the temperature of the cells. The highest yield of $[NaK]^+$ ion signal is obtained at cell temperatures of $180 - 200\,°C$ for Na, and $140 - 150\,°C$ for K. These doping conditions are close to those required for maximum likelihood of single atom doping. Consequently, only low amounts small dopant clusters are observed in the mass spectra. Low concentrations of Rb atoms were obtained from the heated K sample which turned out to be contaminated with Rb. The He droplet beam intensity as well as the alkali doping level is monitored using a beam dump chamber attached to the end of the apparatus which contains a simple surface ionization detector [32].

After passing through a second skimmer, the doped He droplet beam crosses the synchrotron beam at right angles in the center of a VMI-TOF spectrometer [33], capable of detecting multiple ions and electrons in coincidence. The VMI allows one to obtain the kinetic energy distribution of the detected fragments using the MEVELER inversion method [34]. Depending on the voltage applied to the electrodes, electron or ion kinetic energy spectra can be recorded.

## Results and discussion

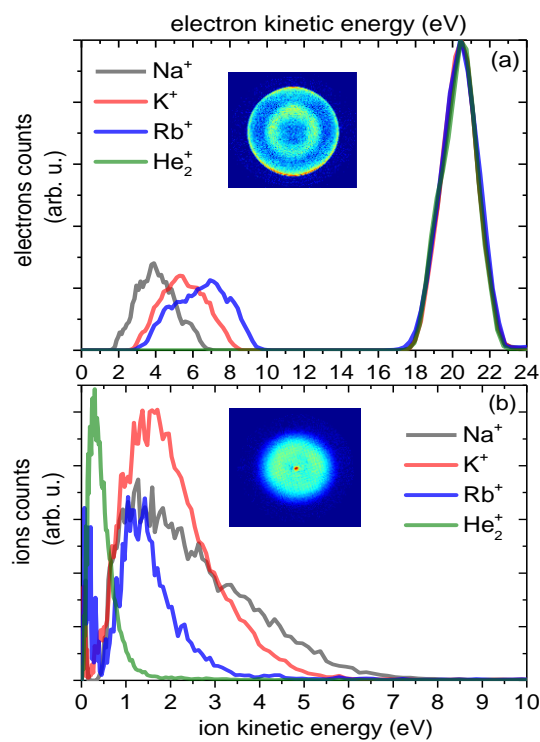

Figure 1: (a) Electron spectra recorded in coincidence with dopant ions and with $He_2^+$ at $h\nu = 45\,eV$. The inset shows the raw electron VMI recorded in coincidence with $K^+$. The peak at 20.4 eV is the photoline of He; ionization of the dopants occurs by radiative charge transfer. The peaks in the range $2 - 9$ eV are due to ETMD of dopant molecules or clusters. (b) Ion kinetic energy distributions measured under identical conditions.

Typical electron spectra measured in coincidence with dopant alkali ions $Ak^+(Na^+, K^+$ and $Rb^+)$ and $He_2^+$ photoions at a photon energy $h\nu = 45\,eV$ are shown in figure 1 (a). These electron spectra are obtained by inverting the raw VMIs of emitted electrons recorded in coincidence with ions as shown in the inset. The raw electron VMI consists of an outer ring indicating the contribution of high kinetic energy electrons (photoelectrons), and an inner diffuse ring indicating a substantial contribution of slow electrons. Two distinct features are therefore apparent in the inverted electron spectrum: a sharp peak at 20.4 eV which corresponds to the photoline of He, and a second, broader feature peaked at about $4 - 7$ eV. The photoline at $20.4\,eV$ is measured in coincidence with the most abundant fragment from ionized He droplet, $He_2^+$, but also in coincidence with $Ak^+$. Note that we do not detect any electrons at energies 0.2-0.6 eV in coincidence with $He_2^+$, which would be created by inelastic energy-loss collisions of the photoelectrons with He atoms in the droplet. Thus, inelastic collisions do not play a significant role at the given experimental conditions. This conclusion is supported by the absence of electrons at energies 12-17 eV in the electron spectra measured in coincidence with the alkali ions, which would be expected if excited He atoms were created in the droplets by inelastic collisions [25]. Ionization of the dopants by interaction with the photoionized

He droplet is conventionally interpreted as due to radiative charge transfer (RCT) [25, 35]. The low energy peaks ($2 - 9$ eV) in the electron spectrum of figure 1 (a) are indicative for ETMD of alkali metal molecules attached to He droplet. For He doped with Mg metal clusters, we had previously observed a low energy peak in the electron spectrum at about 1 eV, which was due to double ionization of the dopants by ETMD [15]. In contrast, the low kinetic energy features seen here (figure 1 a)) are at higher energies due to the lower ionization potentials, $E_i$, of alkali metals. The ratio of peak integrals of each of these ETMD features to the He photoline, which reflects the ETMD efficiency, is about 0.3. However, this number strongly underestimates the ETMD efficiency since Ak monomer dopants which decay only by RCT contribute to the He photoline. Furthermore, this ratio is affected by the probability of detachment of $Ak^+$ from the droplet to reach the detector as free atomic ions, which can strongly vary. In particular, the higher initial kinetic energy of $Ak^+$ produced by ETMD due to Coulomb explosion (see below) tends to enhance their detection and the detection of their correlated electrons because slow ions tend to sink into the droplet where they form strongly bound $Ak^+He_N$ snowball complexes, whereas fast ions can escape the droplet. Figure 1 (b) shows the corresponding kinetic energy distributions of $He_2^+$ and $Ak^+$ dopant obtained by inversion of the raw $Ak^+$ VMIs (see inset). $He_2^+$ cations are ejected out of the He droplet by a non-thermal process in the course of vibrational relaxation [36]; the kinetic energy is about 0.3 eV, which is in agreement with previous measurements [37]. The kinetic energy distributions of $Ak^+$ contain two contributions: A sharp peak at low energy $< 0.4$ eV, which we attribute to RCT from single dopant Ak atoms, and a larger, broad feature peaked around 1.5 eV. The latter matches the kinetic energy of two Ak ions emitted back to back in Coulomb explosion following double ionization of an alkali metal molecule. Double ionization accompanied by the transfer of a single electron, in turn, is another direct manifestation of ETMD. A similar double ionization mechanism caused by energy transfer, termed double ICD, was recently demonstrated. It was found to efficiently create two Ak ions by Coulomb explosion of doubly ionized alkali metal dimer [28].

We note that alkali metal dimers ($Ak_1Ak_2$) are typically formed in their two lowest spin states $^1\Sigma$ or $^3\Sigma$ by aggregation on the He droplet surface. In the $^1\Sigma$ electronic ground state, $Ak_1Ak_2$ is a covalently bound molecule, and, should be considered as a single center in the ETMD reaction. Hence, ETMD(2) is the relevant decay mechanism. In contrast, $Ak_1Ak_2$ dimers in the lowest $^3\Sigma$ state are only weakly bound (~0.05 eV) with the $Ak_1 - Ak_2$ distance being large ($5 - 6$ Å). Therefore, ETMD of this state should be considered as an ETMD(3) process. For the sake of simplicity, in this paper we will refer to double ionization of $Ak_1Ak_2$ by electron transfer to ionized He droplet as ETMD.

Let us now turn to a quantitative analysis of the measured ETMD electron and ion kinetic energy distributions. It is well known that, following photoionization of a He atom inside a He nanodroplets, the He⁺ ion undergoes resonant charge-hoping and eventually localizes by either forming a $He_2^+$ cation, or by inducing charge transfer ionization of a dopant [35, 38]. While the $He_2^+$ formation terminates the charge migration, the $He_2^+$ may still roam about the He droplet on a timescale of ps up to ns owing to the superfluid nature of He nanodroplets. When approaching a dopant dimer, possibly due to steering by long-range forces [39], an electron from one atom of the dopant dimer neutralizes the $He_2^+$ via electron transfer, and the released energy is transferred to the neighboring atom which is then ionized. A possible reason why ETMD takes place after the charge localizes on the $He_2^+$ is that vibrationally excited molecules tend to be ejected toward the droplet surface [36] where the alkali metal monomer and dimer dopants reside in dimple structures [40]. In any case, the observed energetics of the emitted electron clearly indicate that ETMD proceeds from the fully relaxed $He_2^+$ state rather than from He⁺, as previously found for Mg-doped He droplet. Thus, the overall ETMD mechanism in case of NaK attached to a He droplet proceeds as follows:

$$He_2 + NaK + h\nu \rightarrow He_2^+ + NaK + e_{ph}^- \qquad (1)$$

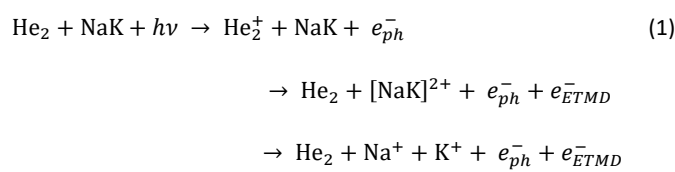

$$\rightarrow He_2 + [NaK]^{2+} + e_{ph}^- + e_{ETMD}^-$$

$$\rightarrow He_2 + Na^+ + K^+ + e_{ph}^- + e_{ETMD}^-$$

Further direct evidence for the ETMD process is obtained by filtering the data for electron-ion-ion triple coincidences. In this way, we obtain time-of-flight ion mass spectra and ion/electron energy distributions for those ionization events where two ions and at least one electron are detected.

When doping He nanodroplets with three species of alkali metals, Na, K, and Rb, dimers form in all possible combinations of species by aggregation of the dopant atoms at the He droplet surface. Figure 2 shows enlarged views of ion-ion coincidence time-of-flight maps for three different heteronuclear alkali dimers (NaK, NaRb, and KRb) recorded at $h\nu = 45$ eV. While the brightness represents the abundance of the events, the shape of the distributions contains information about the fragmentation dynamics [41]. The coincidence maps centred around the masses of two alkali ions show several elongated features. Their negative slopes indicate that fragmentation occurs via Coulomb explosion of the $[Ak_1Ak_2]^{2+}$ dications leading to back-to-back emission of the $Ak_{1,2}^+$. Less intense replicas at larger masses indicate the formation of complexes of an alkali ion with a few attached He atoms. In addition, in each ion-ion coincidence map there is a weaker distribution next to the primary ion pair which mainly originates from the less abundant isotopes ⁴¹K and ⁸⁷Rb. From the primary ion pairs observed in the coincidence maps of figure 2, one can extract the kinetic energy distributions of the coincident electrons as presented in figure 3 (a) that shows the spectrum of the electron measured in triple coincidence (e, ²³Na, ³⁹K) for the NaK case.

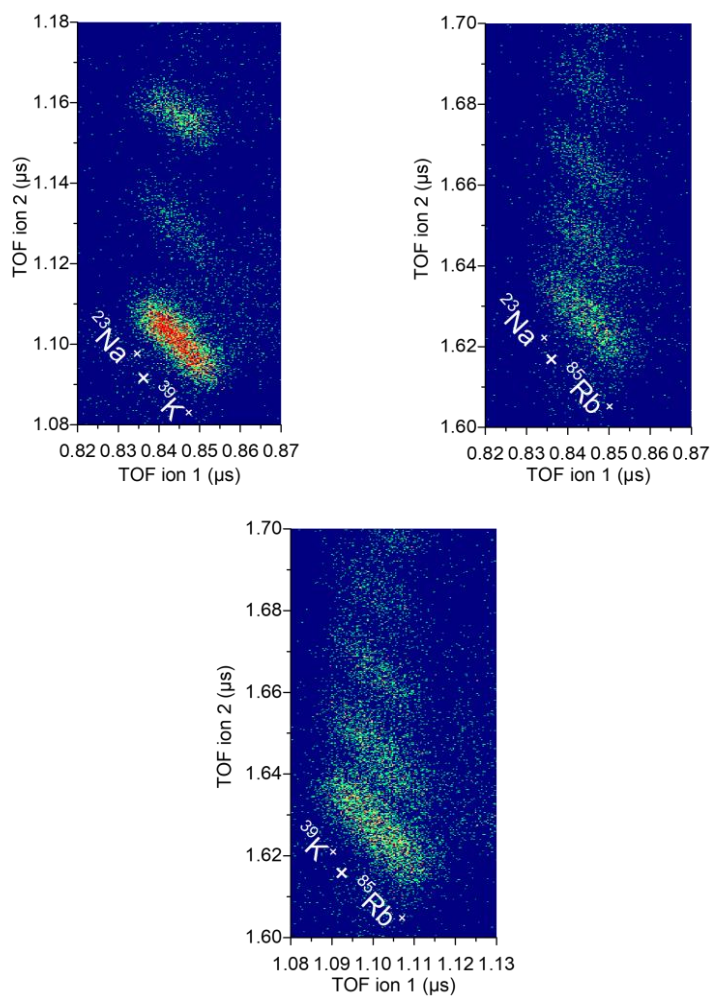

Figure 2: Time-of-flight mass maps recorded in ion-ion-electron triplet coincidence at $h\nu = 45$ eV. Bright regions indicate enhanced signal rates. The anisotropic shapes of the signal distributions indicate Coulomb explosion. The less intense replicas at larger masses are due to abundant isotopes and formation of ion-He complexes of the type $[AkHe]^+$.

Similar to figure 1 (a), the spectrum in figure 3(a) exhibits two features; the one centred at 20.4 eV arises from emission of photoelectrons by photoionization of He atoms within the He droplet followed by electron transfer from the dopant NaK. The other one centred at about 4.3 eV is due to ejection of a second electron from the NaK by ETMD. Figure 3 (b) shows the ion kinetic energy distribution for the two back-to-back emitted ions, i.e. the $^{23}Na^+$ (gray line) and the $^{39}K^+$ (red line), measured in triple coincidence with one electron. The ions have broad kinetic energy distributions centred at about 2.8 eV and 1.7 eV, respectively. The sum of these energies corresponds to the kinetic energy release (KER) of the ion pair in the dicationic state.

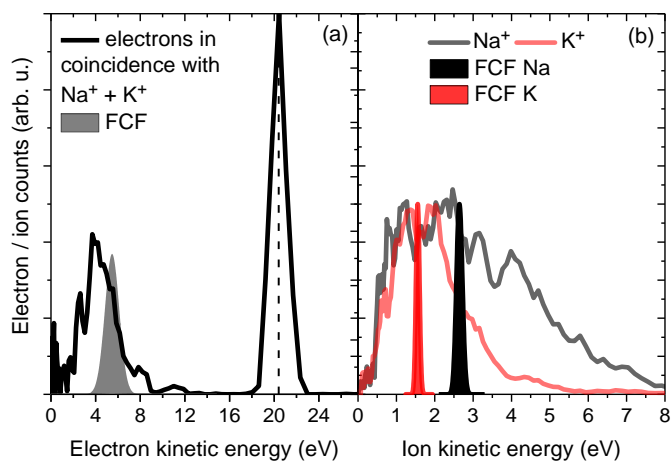

Figure 3: (a) Electron spectrum recorded in triple coincidence with $Na^+$ and $K^+$ at $h\nu = 45$ eV. (b) Ion kinetic energy distributions of the $Na^+$ and $K^+$ fragments detected in triple coincidence. The shaded curves in (a) and (b) represent Franck-Condon profile simulations for the vertical transition from the $^1\Sigma^+$ ground state of NaK to the doubly ionized state $Na^+ + K^+ + 2e^-$.

The double ionization mechanism driven by ETMD is schematically illustrated in figure 4 using the potential energy curves of free NaK in the $^1\Sigma^+$ ground state (black line) [42] and the dicationic state (red line). The dicationic curve was calculated using the Coulomb potential of two positive charges shifted to match the asymptotic ionization energies of the free Na and K atoms.

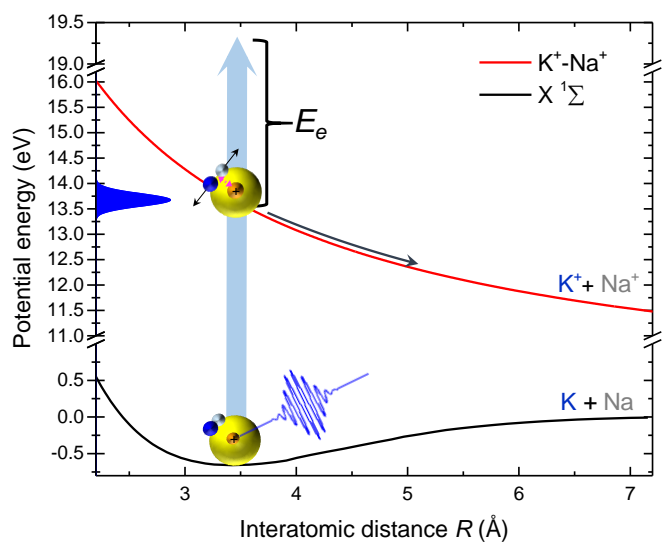

Figure 4: Potential energy scheme illustrating the ETMD process; following photoionization of the He nanodroplets, the charge localizes on a $He_2^+$ ion. This ion is neutralized by ETMD of the NaK dopant molecule in the reaction: $He_2^+ + NaK \rightarrow He_2 + [NaK]^{2+} + e_{ETMD}^-$. The thick vertical arrow indicates the ionization energy of $He_2^+$, $E(He_2^+)$. The kinetic energy $E_e$ of the emitted ETMD electron is given by the excess energy of $E(He_2^+)$ with respect to the vertical double ionization energy of the NaK dimer. The blue filled area indicates the simulated Franck-Condon profile for this transition.

The characteristic electron kinetic energy can be determined by the following equation:

$$E_e = E(\text{He}_2^+) - E_i(\text{Ak}_1) - E_i(\text{Ak}_2) - E_K([\text{Ak}_1\text{Ak}_2]^{2+}) + E_b(\text{Ak}_1\text{Ak}_2) + E_b(\text{He}_2^+\text{-Ak}_1\text{Ak}_2) - E_b(\text{He}_2\text{-}[\text{Ak}_1\text{Ak}_2]^{2+}) \quad (2)$$

where $E(\text{He}_2^+)$ is the energy difference between the ionized and ground states of $\text{He}_2$ at an interatomic distance of 1.1 Å [38, 43, 44], $E_i$ is the ionization potential of the alkali atom $\text{Ak}_{1,2}$, $E_K([\text{Ak}_1\text{Ak}_2]^{2+})$ is the kinetic energy release (KER) of the dicationic state, $E_b(\text{Ak}_1\text{Ak}_2)$ is the binding energy of the alkali dimer in the $^1\Sigma_g^+$ singlet ground state, $E_b(\text{He}_2^+\text{-Ak}_1\text{Ak}_2)$ is the binding energy of $\text{He}_2^+$ to the alkali dimer in the entrance channel, and $E_b(\text{He}_2\text{-}[\text{Ak}_1\text{Ak}_2]^{2+})$ is the binding energy of $\text{He}_2$ to the dicationic alkali dimer in the exit channel. The estimated values of $E_e$ and the other terms in equation (2) are given in Table 1 of the supplementary material. Overall, the estimated electron energy is in good agreement with the experimental data.

To assess the conjecture that the emission of electrons and ions are mainly determined by ETMD of the $\text{He}_2^+$-NaK system, we performed Franck-Condon factor (FCF) simulations (see Annex 1 for more details) of the ion and electron kinetic energy distributions assuming vertical transitions between the potential energy curves shown in figure 4. The shaded grey curve in 3(a) is the result of a convolution of the FCF profiles simulated for the transition out of the $^1\Sigma^+$ ground state of the NaK into the doubly ionized state $\text{Na}^+ - \text{K}^+$ and for the transition from the $\text{He}_2^+$ ionic state into the $\text{He}_2$ ground state which initiates the reaction. A 5 % finite resolution of the VMI spectrometer which may add broadening to the experimental spectra is not included in the convolution of these FCF simulations. The FCF profile is peaked at about 1.2 eV higher in energy as compared to the experimental peak at kinetic energy of about 4.3 eV. This mismatch of kinetic energy between experiment and simulation is attributed to inaccuracies of the estimated energy terms in table 1, and possibly to the population of excitonic satellites in larger Ak clusters which can be formed by electron impact of the ETMD electrons. The results of the FCF simulations for the ion kinetic energy distributions are shown as shaded black peak for $^{23}\text{Na}^+$ and red peak for $^{39}\text{K}^+$ in figure 3 (b). The agreement with the measured values is good with respect to the peak maxima, but drastically underestimates the width. This could be due to both physical and experimental reasons. Firstly, the FCF simulations assume an isolated alkali dimer, not accounting for the complex helium environment, therefore, broadening of the experimental distribution is possibly due to collisions of the emitted ions with the surrounding He atoms in the droplet [45]. Also, perturbations of the initial and final ionic states by the dopant-He droplet interactions which are not taken into account by the FCF simulations may add to the broadening. Additionally, the FCF simulations naturally assume momentum conservation between the two ionic fragments. Experimentally, we plot the ion kinetic energies without applying momentum conservation for each individual event. As such, the distributions plotted in figure 3(b) are the average kinetic energy of the ions and not directly correlated, which can lead to broadened distributions.

Furthermore, depending on the configuration of the initial state, ETMD proceeds at different internuclear distances of the NaK-$\text{He}_2$ droplet system resulting in a broadened energy distribution of the fragmented ions.

Similar to the case of the $\text{He}_2^+$-NaK system, we also experimentally detect ETMD in other pure and mixed $\text{Ak}_1\text{Ak}_2$ dimers attached to He nanodroplets. Figures 5 and 6 show, respectively, the electron and ion kinetic energy distributions of all homonuclear ($\text{Na}_2, \text{K}_2, \text{Rb}_2$) and heteronuclear (NaK, NaRb, KRb) $\text{Ak}_1\text{Ak}_2$ that can form from Na, K, and Rb dopant atoms attached to the surface of He droplet.

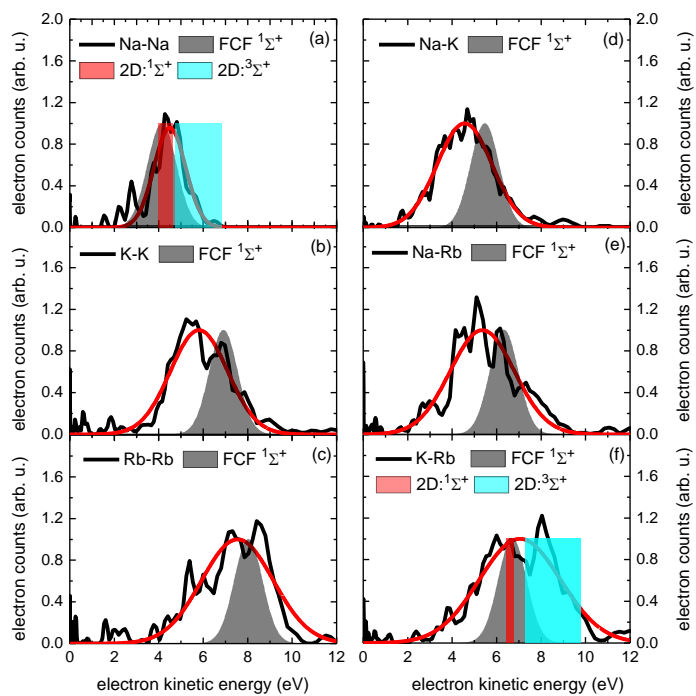

Figure 5: Electron spectra recorded in triple coincidence for homonuclear alkali dimers (left column), and for heteronuclear dimers (right column) attached to He nanodroplets. Each electron spectrum is fitted with a Gaussian function (solid curve) to determine the position of the ETMD peak. The filled curves show the expected electron energies from the FCF simulations. The filled rectangles illustrate the results from high-level *ab initio* calculations using a 2D cut of the potential energy surface.

Owing to the similar electronic structure of all Ak atoms, their electron spectra (figure 5) exhibit similar features with a noticeable energy shift to higher energies due to the decrease in ionization potential from lighter (Na) to heavier (Rb) atoms. The experimental and simulated electron energies are in reasonable agreement. The same as for NaK, the estimated values of $E_e$ in Table 1 (See supplementary material I) for the homonuclear ($\text{Na}_2, \text{K}_2, \text{Rb}_2$) and heteronuclear (NaRb, KRb) $\text{Ak}_1\text{Ak}_2$ also agree reasonably well with the observed peak positions in figure 5 (a), (b), (c) and 5 (e), (f), respectively. Thus, we attribute the low-kinetic energy electron features in figure 1 to ETMD of the dopant alkali dimers.

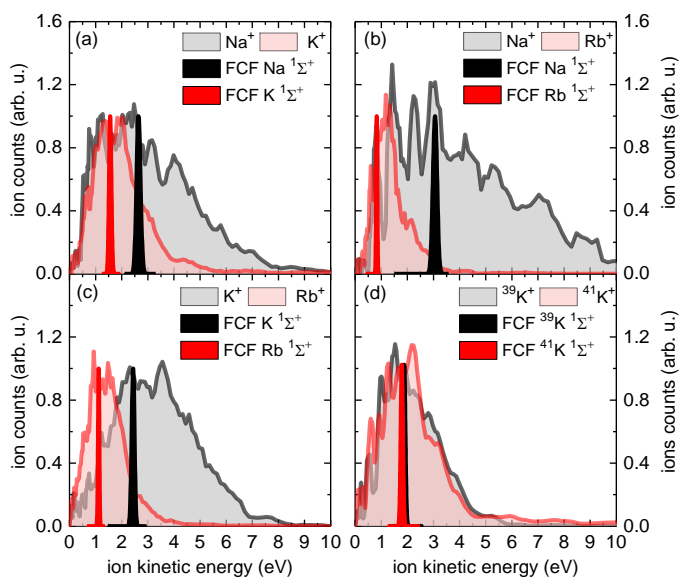

Figure 6: Ion kinetic energy distributions measured in triple coincidence for the heteronuclear alkali metal dimers NaK, NaRb and KRb), and also for $^{39}$K$^{41}$K. The filled curves show the expected ion kinetic energies from the FCF simulation.

Figure 6 shows the corresponding ion kinetic energy distributions for the mixed Ak dimers (panel (a), (b) and (c)) as well as for the potassium isotopologue $^{39}$K $^{41}$K (panel (d)). These measured ion spectra are reasonably well reproduced by the FCF simulations based on the assumption that these molecules are predominantly prepared in their $^1\Sigma^+$ ground states, as shown in figure 6. For the homonuclear alkali dimers, the ion spectra are not reported here because they are subjected to distortions caused by the finite dead time of the detector.

To understand the contribution of different electronic states of the dimers to the observed peaks in the electron spectra, we also carried out *ab initio* calculations of the ETMD electron spectra for He-Na$_2$ and He-KRb (for details see supplementary material II). The computation was done using the following approximations. First, the 3D potential energy surfaces (PES) were reduced to 2D cuts. For the $^1\Sigma^+$ state, where the bonding between the alkali is strong, we assume that the alkali – alkali distance changes little during nuclear dynamics, and we keep the respective coordinate constant at the value in the ground state equilibrium distance. The coordinates, which are being varied, are the distance between the He and the center-of-mass (CoM) of the dimer, and the angle, $\theta$, between the dimer's axis and the line connecting its CoM to He. For the $^3\Sigma$ state, where alkali – alkali is very weakly bound, we assume that there is little motion of He$^+$ around the dimer, so that the angle coordinate is kept at 90°, while the alkali bond distance and the distance between He and the CoM of the dimer are varied.

The spectra were computed by first finding on the He$^+$-Ak$_1$Ak$_2$ PES all points of energy $E$, where $E$ is the energy at the equilibrium nuclear configuration of the neutral cluster. The kinetic energy of the nuclei at these points is zero, and they correspond to the classical turning points in a one dimensional potential. The decay occurs mostly at such turning points where the distance between He$^+$ and the alkali atoms is the smallest, since it is at this point where the ETMD decay time is the shortest [30]. For this subset of turning points we compute the energy of the ETMD electron as the energy difference between the decaying and final HeAk$_1^+$Ak$_2^+$ PES at the respective geometries. To account for the recombination energy difference, $E(\text{He}^+) - E(\text{He}_2^+)$, between He$^+$ and He$_2^+$ at an equilibrium distance of 1.1 Å, we shifted the theoretical spectra by 4.82 eV towards smaller energies.

The results are shown in figures 5(a) and 5(f). The theoretical spectra for both dimers overlap with the higher energy part of the experimental spectra. The peak due to the ETMD with $^1\Sigma$ state of the dimer appears at lower energy than the $^3\Sigma$ peak and overlaps with the FCF peak. The $^1\Sigma$ peak is narrower than the respective $^3\Sigma$ peak. This is due to a much weaker alkali – alkali bond in the latter state so that there is a larger variation in the distances between the alkali ions during the decay and in the final state. The good correspondence between experiment and theory indicates again that the active particle is indeed He$_2^+$.

The theoretical spectrum does not explain the peaks at energies: $< 2.5$ eV for Na$_2$, and $< 6$ eV for KRb. These peaks appear as a series commencing $1 - 2$ eV below the respective $^1\Sigma$ state and proceeding with diminishing intensities towards 0 eV. A possible explanation is as we mentioned, that these peaks are due to the population of ionization satellites in larger alkali clusters which can be formed on the droplet surface. For example, in the case of Na$_3$ one might expect that some Na$^+$Na$^+$Na$^*$ states can be produced in ETMD. Similar situation can be expected for the larger KRb clusters. The energies of these satellites are comparable with the excitation energies in the respective alkali atoms. Indeed the values of the lowest excitations of 2.1 eV for Na, and 1.6 eV for K and Rb match well the gaps between the highest of the unexplained peaks and the theoretical $^1\Sigma$ peak in the respective spectra.

Finally, we assess the importance of ETMD versus ionization by RCT by measuring the kinetic energies of [Ak$_1$Ak$_2$]$^+$ ions at $h\nu = 45$ eV, see figure 7 for the case of K. RCT between K dimers or larger K clusters and the He$^+$ or He$_2^+$ photoion would be observable either as unfragmented K$_2^+$, K$_n^+$ or as K$^+$ ionic fragments generated by dissociation. In the latter case, the K$^+$ energy would be expected around 0.6 eV [46]. However, the K$^+$ spectrum shown in figure 1 (b) does not contain any clear feature at that energy. This indicates that for K$_2$ attached to He nanodroplets, RCT is efficiently quenched by ETMD and therefore the ion kinetic energy spectrum is dominated by K$^+$ + K$^+$ Coulomb explosion. Moreover, even for small K clusters attached to He nanodroplets, we find that ETMD is the dominant ionization process through photoionized He droplet. Figure 7 (b) shows the ion kinetic energy distribution of atomic K$^+$ and K$_2^+$ measured in ion-ion-electron triple coincidence. These signals are due to double ionization of K$_3$ followed by Coulomb explosion of K$^+$ and K$_2^+$ fragments, which is indicative of ETMD. Interestingly, the $K_2^+$ kinetic energy distribution measured in coincidence with the photoelectron only, shown in figure 7 (a), nearly has the same shape as the $K_2^+$ energy distribution of figure 7 (b). This indicates that overall, $K_2^+$ is predominantly created as a fragment from doubly ionized K clusters by ETMD. The lack of a sharp low-energy component close to 0 eV in the $K_2^+$ kinetic energy distribution, which would be expected for $K_2^+$ created by RCT, supports our finding that charge transfer ionization of $K_2$ predominantly proceeds by ETMD thereby generating energetic $K^+$ by Coulomb explosion. This conclusion is in agreement with theory [14], which predicts that the ETMD rate in single Mg atoms doped He clusters is 3 order of magnitude larger than the one of RCT.

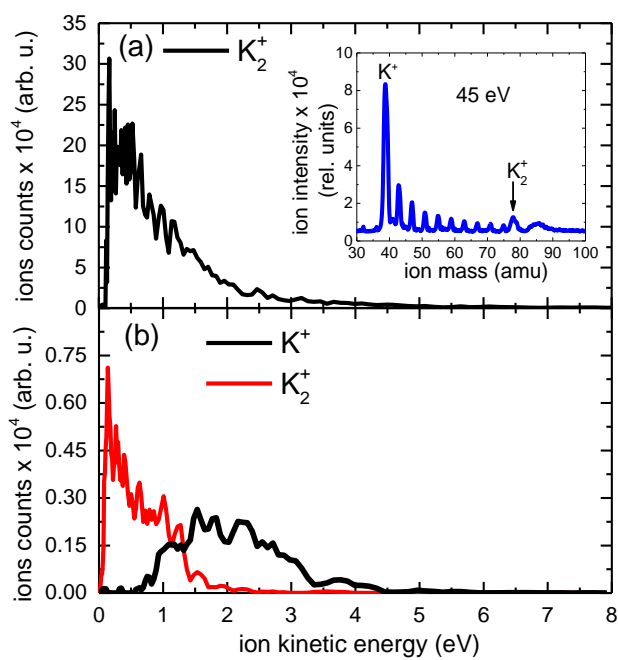

Figure 7: (a) $K_2^+$ ion spectra recorded in ion-electron coincidence at $h\nu = 45$ eV. The inset shows the mass spectrum of He droplet doped with K atoms where the abundance of $K^+$ and $K_2^+$ are clearly visible. (b) Kinetic energy distributions for $K^+$ and $K_2^+$ measured in ion-ion-electron triple coincidence at $h\nu = 45$ eV.

## Conclusions

We have reported the first experimental observation of ETMD for the model system of alkali metal dimers formed on the surface of He nanodroplets. This decay channel is found to be largely dominant in proportion over radiative charge transfer in alkali metal molecules and small clusters. The recorded electron/ion kinetic energy distributions are in good agreement with simulations based on the vertical transition between the initial state – a $He_2^+$ cation interacting with a $Ak_1Ak_2$ alkali dimer in its ground state, and the final state – neutral He, a doubly ionized dopant $[Ak_1Ak_2]^{2+}$, and an electron, when taking the binding energies of the initial and the final states to the He droplet into account. The reasonable agreement between the measured electron/ion spectra and the calculation based on the $^1\Sigma^+$ ground state of the alkali metal dimers reveal the first direct evidence of ETMD(2) in a heterogeneous molecular system. These experiments benefit from the droplet surface location of the alkali dimers which facilitates the detection of free dissociation products generated by ETMD. The sharp non-zero ETMD electrons observed in this work can be considered as an effective primary source of several resonant processes in a surrounding environment such as dissociative electron attachment – an important mechanism causing radiation damage in biological matter [1-3]. For molecular dopants embedded inside the droplet, ETMD may be even more efficient due to the shorter range of the initial state, but detection of the products may be hindered by scattering of electrons and ions at the He shell surrounding the dopant molecule.

## Author contributions

## Conflicts of interest

There are no conflicts to declare.

## Acknowledgements

M.M and L.B.L acknowledge financial support by Deutsche Forschungsgemeinschaft (projects MU 2347/10-1 and BE 6788/1-1:1) and by the Carlsberg Foundation. MM visits IIT Madras and acknowledges support from the SPARC programme (India). S.R.K. thanks Max Planck Society and D.S.T., Govt. of India, for support. K.G gratefully acknowledges the financial support by the European Research Council (ERC) (Advanced Investigator Grant No. 692657). A.C.L. gratefully acknowledges the support by Carl-Zeiss-Stiftung.

## Notes and references

## Annex 1

## FCF simulation of the ion and electron kinetic energy distributions of alkali dimers created by ETMD on He nanodroplets.

The simulation of the kinetic energy distributions of the alkali ions $Ak^+$ generated by ETMD and the spectra of the ETMD electrons is based on the Franck-Condon factor (FCF) for the vertical bound-continuum transition from the $Ak_1Ak_2$ dopant ground state $^1\Sigma^+$ into the doubly ionized state $[Ak_1Ak_2]^{2+}$ (figure 4):

$$FCF(V) = \left| \int_{-\infty}^{\infty} \Psi_{X,v=0}(R)\, \Psi_{Ak^{2+}}^V(R)\, dR \right|^2$$

Here, $\Psi_{X,v=0}$ is the vibrational wave function of the neutral $Ak_1Ak_2$ dimer, $\Psi_{Ak^{2+}}^V$ is the continuum wave function of the dissociating $[Ak_1Ak_2]^{2+}$ at potential energy $V$, and $R$ is the interatomic distance of the $Ak_1Ak_2$ dimer. Owing to the low temperature of the He nanodroplets ( 0.37 K), we assume $Ak_1Ak_2$ to be initially prepared in the vibrational ground state ($v=0$). Since alkali atoms and small clusters are attached to He nanodroplets in weakly bound surface states [47], the perturbation of the intramolecular potential energy curves by the He droplet is neglected in the simulation.

FCF(*V*) is calculated numerically using the program BCONT2.2 [48]. From FCF(*V*), we obtain the kinetic energy distribution of the $Ak_1^+$ and $Ak_2^+$ ionic fragments, $P_{Ak_1^+}(KE)$, by linear transformation of the argument,

$$P_{Ak_1^+}(KE) = FCF\left[\frac{m_{Ak_2}}{m_{Ak_1}+m_{Ak_2}}(V - E(v=0) - IP(Ak_1) - IP(Ak_2))\right],$$

where $m_{Ak_1}$ and $m_{Ak_2}$ are the respective masses of the $Ak_1$ and $Ak_2$ atoms of the $Ak_1Ak_2$ dimer. Here, $E(v=0)$ is the energy of the $v=0$ lowest vibrational level in the $^1\Sigma^+$ state potential of the $Ak_1Ak_2$ dimer. $I_p(Ak_1)$ and $I_p(Ak_2)$ denote the ionization energies of the $Ak_1$ and $Ak_2$ atoms, respectively. The energy distribution of the ETMD electron, $P_{e_{ETMD}}$, is obtained from

$$P_{e_{ETMD}}(E_{e_{ETMD}}) = FCF[E - V].$$

Here, $E = E(He_2^+) + E_b(He_2^+\text{-}Ak_1Ak_2) - E_b(He_2\text{-}[Ak_1Ak_2]^{2+})$ where $E(He_2^+) \approx 19.77\ eV$ denotes the difference in energy between the ionized and ground states of $He_2$ at the $He_2^+$ equilibrium distance of 1.1 Å, and $E_b(He_2^+\text{-}Ak_1Ak_2) - E_b(He_2\text{-}[Ak_1Ak_2]^{2+})$ is the difference between the binding energies of the $He_2^+\text{-}Ak_1Ak_2$ system in the entrance and exit channels (See Table 1 in supplementary material I for the values of the binding energies).